\documentclass[aps,prl,twocolumn,showpacs,preprintnumbers,amsmath,amssymb]{revtex4-1}
\usepackage{graphicx}
\usepackage{dcolumn}
\usepackage{subfigure}
\usepackage{wrapfig}
\usepackage{cancel}
\usepackage{color}
\usepackage{bm}

\begin{document}

\def\k{{\bf k}}
\def\rr{{\bf r}}
\def\q{{\bf q}}

\title{\bf $d$-wave pairing from spin fluctuations in the K$_x$Fe$_{2-y}$Se$_2$ superconductors}

\author{T. A. Maier$^1$, S. Graser$^2$, P. J. Hirschfeld$^3$, and D. J. Scalapino$^4$}

\affiliation{
$^1$ Center for Nanophase Materials
Sciences and Computer Science and Mathematics Division, Oak Ridge
National Laboratory, Oak Ridge, TN 37831-6494 \\
$^2$ Center for Electronic Correlations and Magnetism, Institute of Physics,
University of Augsburg, D-86135 Augsburg, Germany \\
$^3$Department of Physics, University of Florida,
Gainesville, FL 32611, USA \\
$^4$ Department of Physics, University of California, Santa
Barbara, CA 93106-9530 USA}

\date{\today}

\begin{abstract}
Angle-resolved photoemission spectroscopy measurements on the recently discovered superconductors in
the KFe$_2$Se$_2$ family with critical temperatures up to $\sim 33K$ suggest that no Fermi pockets of hole character
centered on the $\Gamma$ point of the Brillouin zone are present, in contrast to all other known ferropnictide
and ferrochalcogenide superconductors.  Using a fluctuation exchange approximation and a 5-orbital tight-binding
description of the band structure, we calculate the effective pairing interaction.
We find that the pairing state in this system is  most likely to have $d$-wave symmetry due to pair scattering
between the remaining electron Fermi pockets at wave vector ${\bf q}\sim (\pi,\pi)$, but without any symmetry-imposed
nodes for the given Fermi surface.  We propose experimental tests of this result, including the form of the resonance
spectrum probed by inelastic neutron scattering.
\end{abstract}


\maketitle

{\it Introduction.}
Recently a new family of Fe-chalcogenide superconductors
A$_x$Fe$_{2-y}$Se$_2$ (A=K,Ca) with $T_c \propto$ 30K has been discovered~\cite{ref:Guo}.
These compounds are heavily electron doped, such that there are only electron
Fermi surface pockets, according to angle-resolved photoemission (ARPES) studies~\cite{ref:Qian}.
The usual argument \cite{ref:Mazin,ref:Kuroki,ref:NJP} leading to the most popular ``$s_\pm$" gap structure in the Fe-based
superconductors requires a $\Gamma$-centered pocket to enhance spin fluctuation
pairing with wave vector ${\bf Q}\sim (\pi,0)$ in the unfolded (1-Fe) Brillouin zone.
In the absence of these hole pockets,
which were present in the previously studied Fe-based superconductors,
a gap of the $s_\pm$ type is unlikely. Thus the question of the pairing mechanism
and the structure of the gap in these materials remains open.
One possibility is that the pairing interaction is  associated with
the exchange of spin-fluctuations, as considered for the older materials \cite{ref:Kuroki,ref:Ishida,ref:NJP,ref:fRG,ref:FLEX},
but that the effective interaction peaks at  a wave vector
$Q=(\pi,\pi)$  rather than at $(\pi,0)$. In this case one would expect that the gap would have
B$_{1g}$ ($d$-wave) symmetry, changing sign between the $(\pi,0)$ and $(0,\pi)$
electron Fermi surfaces.  Since there are no portions of the Fermi surface
along the $(\pi,\pi)$ direction in the Brillouin zone, there is no symmetry reason in a 2D $d$-wave gap to have nodes.

The situation reported by ARPES presents an interesting new theoretical challenge
in these systems, namely the calculation of the possible types of superconductivity arising in the
presence of small pockets of only one type of carrier. The proximity of
$d$-wave pairing to the dominant $s_\pm$ pairing channel in the pnictides generally was discussed in earlier spin-fluctuation theories\cite{ref:Kuroki,ref:NJP}, and its likelihood in the situation with pockets of one type was mentioned briefly in  earlier work \cite{ref:Kuroki,ref:Kemper},
but was not explored seriously.   Recently, Thomale et al.\cite{ref:Thomale11} showed that
$d$-wave pairing was likely in the 3K superconductor KFe$_2$As$_2$, which is believed to possess only hole pockets.  The chalcogenide analog
K$_x$Fe$_{2-y}$Se$_2$ is of considerable interest not only because the opposite situation obtains, but because the critical temperature is an order of magnitude higher.

 If superconductivity is possible with only one type of
pocket, there are in addition two qualitatively different  situations to address: one in which the hole
and electron bands overlap in energy, but only one pocket type is present due to
heavy doping. In addition, a new type of situation is suggested
by ARPES, one in which an energy gap exists between two bands.  Doping may then
lead to a transition between two different symmetry superconducting states, or from  a superconducting state with one  symmetry, to an insulating state, and then to another symmetry superconductor.
  These different types of gap symmetry transitions as a function
of doping will be important to explore.

The main results of our paper are obtained from a fluctuation exchange
RPA calculation for a five orbital tight binding model based on an LDA
calculation for these Fe chalcogen compounds. Within this model, using typical
interaction parameters, we determine the structure of the gap.  We then discuss  ways in which the nodeless d-wave
state might be distinguished from ordinary s-wave super-
conductivity, focussing particularly on
the neutron scattering response that would
be expected in such states.

{\it Model.}
In the following, we consider a general two-body onsite interaction Hamiltonian
\begin{eqnarray}
H & = & H_{0}+\bar{U}\sum_{i,\ell}n_{i\ell\uparrow}n_{i\ell\downarrow}+\bar{U}'\sum_{i,\ell'<\ell}n_{i\ell}n_{i\ell'} \nonumber\\
 &  & +\bar{J}\sum_{i,\ell'<\ell}\sum_{\sigma,\sigma'}c_{i\ell\sigma}^{\dagger}c_{i\ell'\sigma'}^{\dagger}c_{i\ell\sigma'}c_{i\ell'\sigma}\\
 &  & +\bar{J}'\sum_{i,\ell'\neq\ell}c_{i\ell\uparrow}^{\dagger}c_{i\ell\downarrow}^{\dagger}c_{i\ell'\downarrow}c_{i\ell'\uparrow} \nonumber
\label{H}
\end{eqnarray}
where the interaction parameters $\bar{U}$, $\bar{U}'$, $\bar{J}$, $\bar{J}'$ are used in the notation of Kuroki {\it et al.}~\cite{ref:Kuroki}.
The tight-binding Hamiltonian $H_0$ is fitted to the full DFT band structure of the parent compound KFe$_2$Se$_2$, calculated within a
plane wave basis set with ultrasoft pseudopotentials using the tools of the quantum espresso package~\cite{ref:QE}. A Wannier projection~\cite{ref:Marzari}
onto a 10-orbital Fe $d$ basis allows the determination of the position and the orbital composition of the energy bands that can
then be fitted by a reduced 5-orbital tight-binding model similar to the one found for the isostructural BaFe$_2$As$_2$~\cite{ref:Ba122}.

\begin{figure}
\includegraphics[width=\columnwidth]{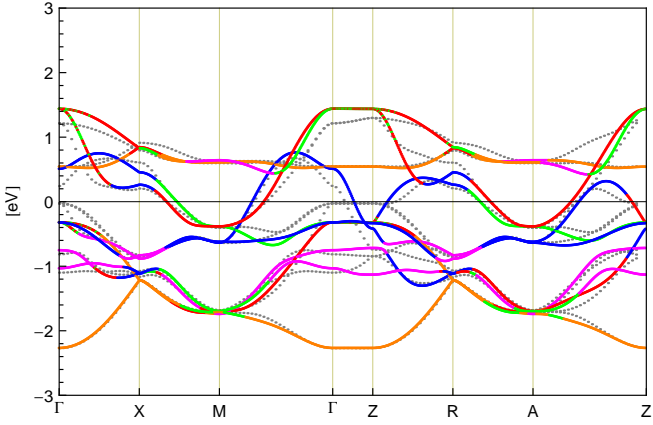}
\caption{(Color online) The 5-orbital fit to the LDA band structure with colors indicating the majority orbital character
(red=$d_{xz}$, green=$d_{yz}$, blue=$d_{xy}$, orange=$d_{x^2-y^2}$, and magenta=$d_{3z^2-r^2}$).
The gray points indicate the 10-orbital Wannier fit to the full DFT band structure.
The splitting of the $d_{xz}/d_{yz}$ and the $d_{xy}$ bands at $\Gamma$ has been enlarged to
remove the hole pockets.
}
\label{fig:bandstructure}
\end{figure}

\begin{figure}
\includegraphics[width=0.6\columnwidth]{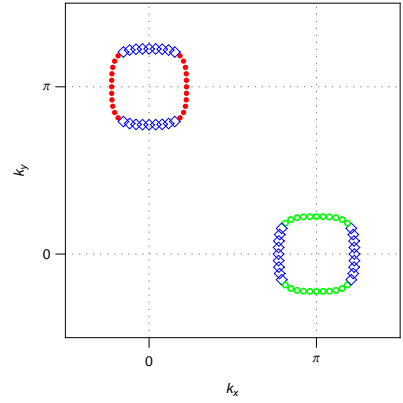}
\caption{(Color online) The Fermi surface of the 5-orbital tight-binding fit with $\mu=E_F-0.2$.
The colors represent the major orbital character of the Fermi surface with the same
color code as in in Fig.~\ref{fig:bandstructure}.}
\label{fig:fermisurface}
\end{figure}

Considering the recent ARPES resuls on K$_{0.8}$Fe$_{1.7}$Se$_2$
reported by T.~Qian {\it et al.}~\cite{ref:Qian} we have artificially enhanced the splitting between the two
$d_{xz}/d_{yz}$ bands and the two $d_{xy}$ bands at the $\Gamma$ point of the backfolded Brillouin zone that is controlled by the nearest neighbor hopping
$t_x(d_{xz},d_{xz})$ and $t_x(d_{xy},d_{xy})$, respectively. This allows us to push the hole pockets below the Fermi level without changing
the orbital character of the respective bands (see Fig.~\ref{fig:bandstructure}).
The effect of a simultaneous downward shift of the hole bands together with an upward shift of the electron bands
reducing the difference between the bottom of the electron bands and the top of the hole bands
can be explained if one accounts for the dominant interband interactions present in the pnictides~\cite{ref:Ortenzi}.
We have also adjusted the chemical potential to account for the reduced electron doping of K$_x$Fe$_{2-y}$Se$_2$ with
$x=0.8$ and $y=0.3$ (0.1 electrons per Fe) compared to the parent compound with $x=1$ and $y=0$ (0.5 electrons per Fe).
In Fig.~\ref{fig:fermisurface} the Fermi surface for $\mu=E_F-0.2$ is shown with a color encoding of the majority orbital character.  Note that the square
Fermi surface pockets found here allow for the possibility of nesting at vectors
away from $(\pi,\pi)$.

To determine the pairing symmetry arising from a spin fluctuation exchange picture, we define the
following scattering vertex $\Gamma(\k,\k')$ in the singlet channel,
\begin{eqnarray}
{\Gamma}_{ij} (\k,\k') & = & \mathrm{Re}\sum_{\ell_1\ell_2\ell_3\ell_4} a_{\nu_i}^{\ell_2,*}(\k) a_{\nu_i}^{\ell_3,*}(-\k) \\
&&\times \left[{\Gamma}_{\ell_1\ell_2\ell_3\ell_4} (\k,\k',\omega=0) \right] a_{\nu_j}^{\ell_1}(\k')
a_{\nu_j}^{\ell_4}(-\k')\nonumber
\label{eq:Gam_ij}
\end{eqnarray}
Here the momenta $\k$ and $\k'$ are restricted to the electron pockets $\k \in C_i$  and $\k' \in C_j$ where
$i$ and $j$ label either the $\beta_1$ or the $\beta_2$ Fermi surface, and $a^{l}_{\nu}(\k)$ are the orbital-band matrix-elements. The
orbital vertex functions $\Gamma_{\ell_1\ell_2\ell_3\ell_4}$ represent the
particle-particle scattering of electrons in orbitals $\ell_1,\ell_4$ into $\ell_2,\ell_3$
and are given by
\begin{eqnarray}
&&{\Gamma}_{\ell_1\ell_2\ell_3\ell_4} (\k,\k',\omega) = \left[\frac{3}{2} \bar U^s
\chi_1^{\rm RPA}  (\k-\k',\omega) \bar U^s + \nonumber \right.\,~~~~~~\,\\
&&\,~~~~~\left.
 \frac{1}{2} \bar  U^s
 - \frac{1}{2}\bar U^c  \chi_0^{\rm RPA}  (\k-\k',\omega)
\bar U^c + \frac{1}{2} \bar U^c \right]_{\ell_1\ell_2\ell_3\ell_4}
\label{eq:fullGamma}
\end{eqnarray}
The interaction matrices $\bar U^s$ and $\bar U^c$ in orbital space
are built from linear combinations of the interaction parameters. Their explicit form
can be found e.g. in Ref.~\cite{ref:Kemper}.
Here $\chi_1^{\rm RPA}$ and $\chi_0^{\rm RPA}$ denote the spin-fluctuation
contribution and the orbital-fluctuation contribution to the RPA susceptibility, respectively.

The symmetry function $g(\k)$ of the pairing state can then be found by
solving an eigenvalue problem of the form
\begin{equation}
- \sum_j \oint_{C_j} \frac{d \k_\parallel'}{2\pi v_F(\k_\parallel')} \Gamma_{ij} (\k,\k') g (\k')
  = \lambda g(\k)
\end{equation}
where the eigenfunction $g(\k)$ corresponding to the largest eigenvalue $\lambda$
gives the leading pair instability of the system.

\begin{figure}
\includegraphics[width=1.0\columnwidth]{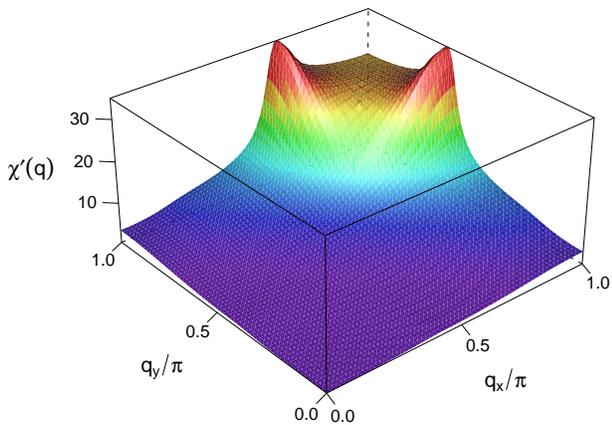}
\caption{(Color online) The real part of the RPA enhanced static susceptibility
for $U=0.96$, $J=U/4$ in the normal state showing a broad peak at $\q = (\pi,\pi)$.}
\label{fig:susceptibility}
\end{figure}

In the following we parameterize the superconducting gap in low-order harmonics
and calculate the susceptibility in the symmetry-broken state as~\cite{ref:ScalapinoMaier,ref:Maieretal}
\begin{eqnarray}
 \chi_{rstu}^0 (q) & = & -\frac{1}{2} \sum_{k,\mu\nu} M_{rstu}^{\mu\nu} (\k,\q) \\
 & & \times \{ G^{\mu}(k+q) G^{\nu} (k) + F^{\mu}(-k-q) F^{\nu} (k) \} \nonumber
\end{eqnarray}
where the generalized momenta $q=(\q,\omega_m)$ and $k=(\k,\omega_n)$ comprise both the
momenta and the Matsubara frequencies. The normal and anomalous Green's functions are given as
\begin{equation}
 G^\mu(k)= \frac{i \omega_n + \xi_\nu(\k)}{\omega_n^2 + E_\nu^2(\k)}, \;\;\;
 F^\mu(k)= \frac{\Delta(\k)}{\omega_n^2 + E_\nu^2(\k)}
\end{equation}
Here the matrix elements connecting band and orbital space determine
\begin{equation}
 M_{rstu}^{\mu\nu} (\k,\q) = a_\mu^{r,*} (\k+\q) a_\nu^{s} (\k) a_\nu^{t,*} (\k) a_\mu^{u} (\k+\q)
\end{equation}
and the quasiparticle energies for a band $\nu$ are given as $E_\nu(\k) = \sqrt{\xi_\nu^2(\k) + \Delta^2(\k)}$. The inelastic
neutron intensity is proportional to the imaginary part of the spin susceptibility
in the symmetry-broken state
\begin{equation}
 \chi (\q,\omega) = \sum_{rt} \chi_{rrtt}^{\rm RPA} (\q,\omega)
\end{equation}
The multiorbital RPA enhanced spin susceptibility is defined by a Dyson-like equation
\begin{equation}
 \chi_{rstu}^{\rm RPA} (q,\omega) =  \left\{ \chi^0 (q,\omega) \left[1 -\bar U^s \chi^0 (q,\omega) \right]^{-1} \right\}_{rstu}
\end{equation}
with interaction matrices $U^s$ as discussed in Ref.~\cite{ref:Kemper}.

\begin{figure}
\includegraphics[width=1.1\columnwidth]{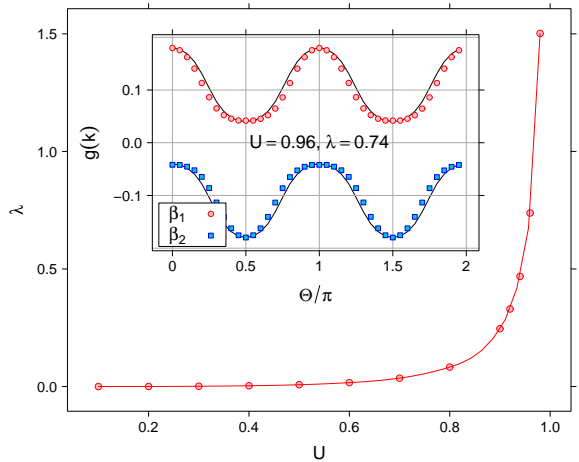}
\caption{The leading $d$-wave eigenvalue as a function of the intra-orbital
interaction strength $U$.  Insert: the symmetry function $g(\k)$ for $U=0.96$, $J=U/4$ along the electron pockets $\beta_1$ (red) and $\beta_2$ (blue) as a function of the winding angle $\theta$ parameterizing the Fermi vector $\k$ relative to the
center of each pocket. A parameterized fit of $g(\k)$ used in the neutron scattering calculation is shown as the solid line.}
\label{fig:lambda}
\end{figure}
{\it Results.} The absence of the hole pocket around $\Gamma$ in the unfolded 1 Fe per unit cell Brillouin zone removes the dominant
$\q=(\pi,0)$ nesting that is characteristic of the LaOFeAs and the BaFe$_2$As$_2$ compounds and is responsible not only for
the stripe-like SDW instability of the undoped parent compounds but also for the sign changing $s$-wave superconducting state as first pointed out by
Mazin {\it et al.}~\cite{ref:Mazin} and later consistently reported by RPA, fRG and FLEX calculations for the doped superconducting
materials~\cite{ref:Kuroki,ref:NJP,ref:fRG,ref:FLEX}. With only the electron pockets present, the real part of the susceptibility
reflecting the nesting properties of the electron pockets around $(\pi,0)$ and $(0,\pi)$ takes a broad
plateau-like maximum around $\q=(\pi,\pi)$ that is bordered by two weak peaks at $\q \approx (\pi,0.625\pi)$
and $\q \approx  (0.625 \pi,\pi)$ resulting from an enhanced scattering between the flat top and bottom parts of the $\beta_{1}$ and $\beta_{2}$ sheets respectively (compare Fig.~\ref{fig:fermisurface}). Here we note that the broad flat maximum corresponding
to the nesting of the two electron pockets is in contrast to the sharp peak features expected from the nesting
of an electron and a hypothetical hole pocket. The  orbital (charge) susceptibility not shown here has no
pronounced momentum space structure and does not approach an instability for the parameters chosen.

In Fig.~\ref{fig:lambda} we show the leading $d$-wave eigenvalue as a function of $U$ for a Hund's rule coupling $J=U/4$,
a pair hopping term $J'=J$ and an inter-orbital Coulomb interaction $U'=U-2J$ where the latter two are fixed by the
local spin-rotational invariance. It reaches an instability around $U=1$ corresponding to an anisotropic but nodeless
superconducting gap on the electron sheets as shown in Fig.~\ref{fig:lambda}.
Due to the $I4/mmm$ symmetry of the crystal, the backfolding of the bands from the effective BZ of the 1 Fe unit cell to the
real BZ of the 2 Fe unit cell leads to two concentric electron pockets of different size around the $M$ point of the backfolded zone.
Due to the origin of the two concentric electron pockets in the backfolded BZ from different positions in the unfolded zone, the
superconducting gap exhibits a phase difference of $\pi$ between them but respects the overall $B_{1g}$ symmetry.

The present calculation was performed using the 1-Fe Brillouin zone representation of the band states, which neglects the hybridization between the two electron pockets backfolded onto one another via the body-centered cubic symmetry operation translation by $(\pi,\pi,\pi)$.  It is interesting to ask how the result will change if this hybridization or spin-orbit coupling is included.  In the simplest case, if the unhybridized pockets do not cross at $k_z=0$ or $\pm \pi$ when backfolded, the hybridization will affect only the states near $k_z=\pm\pi/2$, leading to a
horizontal node if the pairing interaction is relatively $k_z$ independent, as we have found
in other studies \cite{ref:Ba122}.  If the band structure involves crossings at $k_z=0$ or $\pm \pi$, nodes with
some vertical character may be formed on the electron sheets.  In contrast to nodes driven by the dominant point group symmetry microscopic 2D interaction, however, these nodes are a consequence of the crystal space group symmetry and their strength is proportional to the hybridization between the 1-Fe bands.  The density of associated quasiparticle excitations is therefore expected to be weak.  Further calculations in the 2-Fe zone are underway to confirm this.

To calculate the neutron response we parametrize the superconducting gap as $\Delta(\k)=\Delta_{0}g(\k)$ with
\begin{equation}
 g(\k) = (\cos k_x -\cos k_y) + 1.62 (\cos 2k_x -\cos 2k_y)
\end{equation}
This fit is shown as the solid line in the inset of Fig.~\ref{fig:lambda}.
\begin{figure}
\includegraphics[width=1.05\columnwidth]{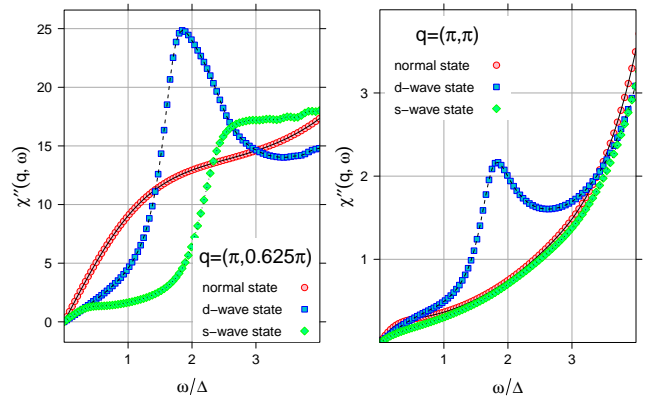}
\caption{(Color online) The imaginary part of the RPA-BCS dynamic spin susceptibility
versus $\omega$ for $\q=(\pi,0.625 \pi)$ (a) and $\q=(\pi,\pi)$ (b) for the normal, d-wave and s-wave states. The interaction parameters were chosen as
$U=0.96$ and $J=U/4$.}
\label{fig:neutronresponse}
\end{figure}
In Fig.~\ref{fig:neutronresponse} we show the imaginary part of the RPA enhanced
susceptibility in the $d$-wave state for a momentum transfer of $\q=(\pi,0.625\pi)$ in (a) and $\q=(\pi,\pi)$ in (b)
in the 1Fe/unit cell Brillouin zone. Here the low energy spectral weight is
suppressed upon the opening of the superconducting gap and is transferred to higher energies.
The pronounced resonance peak around $\omega = 2 \Delta$ appears only for a
relative sign change of the gap on the two electron pockets such that the coherence factor $1-\Delta(\k)\Delta(\k+\q)/(E(\k) E(\k+\q))$ does not vanish.

There are proposals that an s-wave gap may arise from the orbital term in Eq.~\ref{eq:fullGamma} when local Fe phonon modes are included \cite{ref:Saito10,ref:Yanagi10}. In Fig.~\ref{fig:neutronresponse} we have added results for an s-wave gap  taken equal to the average magnitude of the d-wave gap for comparison.  While an s-wave gap could well have anisotropic structure, we expect that the difference in $\chi^{\prime\prime}(\q,w)$ between a B$_{1g}$ and an A$_{1g}$ gap will be significant. In particular if the orbital fluctuations are dominant, the response in the magnetic scattering channel will be further suppressed compared to the d-wave response illustrated in Fig.~\ref{fig:neutronresponse}.

{\it Conclusions.} We have argued, based on an RPA treatment of a generalized multiorbital Hubbard model, that the absence of the $\Gamma$-centered hole pocket in the KFe$_2$Se$_2$
 superconducting materials should lead directly to a strong $d$-wave pairing instability without 
 nodes on the remaining $M$-centered electron pockets.  The appearance of $d$-wave pairing in this family of unconventional superconductors in the limit when only one pocket is present would be strong support for pairing by spin fluctuations in these systems.  It appears to us that the measurement of a peak in the inelastic neutron scattering spectrum near $\pi,\pi$ would be the easiest way to test this prediction.

 Since the inelastic neutron scattering is mostly sensitive to the Fe lattice, it is
possible to distinguish with this technique between the low $\q$ and the $\q=(\pi,\pi)$ scattering in
the unfolded 1Fe/unit cell Brillouin zone, although both signals would be backfolded on the
$\Gamma$ point in the 2Fe/unit cell relevant e.g. for the interpretation of the
angle resolved photoemission results. Therefore the proposed experiment can provide a direct
measurement of the $\q$ vector dominating the repulsive interaction and eventually leading to a
sign change of the superconducting gap on Fermi surface regions connected by $\q$.

In a recent preprint using a functional renormalization group approach, F. Wang {\it et al.} also conclude that the leading pairing instability of a K$_{x}$Fe$_{2-y}$Se$_{2}$ model occurs in the $d_{x^{2}-y^{2}}$ channel \cite{ref:Wang}.

{\it Acknowledgements.} This work is supported by DOE DE-FG02-05ER46236 (PJH)
and the DFG through TRR80 (SG). TAM and DJS acknowledge support from the Center for Nanophase Materials Sciences, which is sponsored at Oak Ridge National Laboratory by the Scientific User Facilities Division, U.S. Department of Energy. All authors are grateful to
the hospitality and the vibrant and inspiring atmosphere at KITP,
where this work was performed.
We would also like to acknowledge fruitful discussions
with A.~Chubukov and I.~Mazin.

\end{document}